\documentclass[twocolumn]{revtex4}

\usepackage{amsmath}    

\newcommand{\Sec}[1]{Sec.~\ref{#1}}
\newcommand{\Fig}[1]{Fig.~\ref{#1}}
\newcommand{\Tab}[1]{Table~\ref{#1}}

\usepackage{graphicx}   
\usepackage{verbatim}   
\usepackage{color}      
\usepackage{hyperref}   
\newcommand{\be}{\begin{equation}}
\newcommand{\ee}{\end{equation}}
\newcommand{\bi}{\begin{itemize}}
\newcommand{\ei}{\end{itemize}}

\newcommand {\unit} [1] {\; \mathrm {#1}}
\newcommand {\cL} {\mathcal{L}}
\newcommand {\cO} {\mathcal{O}}

\DeclareMathOperator{\U}{U}

\begin{document}

\title{Semielastic Dark Matter}
\author{David Krohn}
\affiliation{Department of Physics, Princeton University, Princeton, NJ 08540}
\author{Joshua T. Ruderman}
\affiliation{Department of Physics, Princeton University, Princeton, NJ 08540}
\author{Lian-Tao Wang}
\affiliation{Department of Physics, Princeton University, Princeton, NJ 08540}
\date{\today}
\begin{abstract}
Many models have recently been proposed in which dark matter (DM) couples to Standard Model
fields via a GeV-scale dark sector.  We consider
scenarios of this type where the DM mass, at the electroweak/TeV scale, is generated by the VEV of a singlet which also couples to the Higgs.
Such a setup 
results in a distinct recoil spectrum with both elastic and inelastic components.
We construct an explicit NMSSM-like realization of this setup,
discuss constraints coming from the relic density, and include benchmark points which are consistent with current limits, yet visible at upcoming direct detection experiments.
\end{abstract}
\maketitle

\section{Introduction}
By now, there is overwhelming evidence supporting the existence of particle dark matter (DM)\@.
Many models incorporate DM into proposals for new TeV-scale physics, 
with  recent efforts focusing on the possibility that DM may also interact with a {\it dark sector}, composed of new gauge groups and light (GeV-scale) degrees of freedom.  
These models attribute positron/electron cosmic ray excesses \cite{Cosmics} to the annihilation \cite{ArkaniHamed:2008qn, ArkaniHamed:2008qp, OtherDS, Baumgart:2009tn, Cheung:2009qd} or decay of \cite{OtherDecayDS, JoshDecayDS} of DM via the new GeV-scale states.  In effect, the dark sector serves as a restrictive portal, allowing DM to decay/annihilate to light
leptons, but not into (unobserved) protons and anti-protons.
Dark sectors can also naturally induce an $\cO(100)$ keV mass splitting between the DM states~\cite{ArkaniHamed:2008qn,ArkaniHamed:2008qp}.  
Consequently, if the fields mediating the scattering of DM with atomic nuclei couple off-diagonally to different DM states,
the result is a novel  inelastic  recoil spectrum~\cite{Inelastic} which has been used to explain the DAMA results \cite{Bernabei:2010mq}.  Indeed, regardless of any DAMA signal,  inelastic scattering is a generic consequence of dark sectors which can be probed by upcoming experiments.

The phenomenology summarized above follows from the connection between GeV-scale fields and the Standard Model (SM).  But this is unlikely to be the whole story, because we would like the TeV-scale mass of DM to be related to the scale of electroweak symmetry breaking.  The simplest way to generate the mass of DM is to couple it to a singlet that also couples to the SM-Higgs and receives a TeV-scale VEV~\cite{ArkaniHamed:2008qp, HiggsSDM}.  This naturally allows DM to scatter elastically off of atomic nuclei via the exchange of a Higgs/singlet.
{\it We find that such dark sector models yield a distinct recoil spectrum, with both elastic and inelastic components, visible at the next generation of direct detection experiments such as XENON100.}  

This paper is structured as follows.  In \Sec{sec:semielastic} we introduce this new recoil spectrum and survey its unique features.  In \Sec{sec:model} we construct an explicit NMSSM-like model realizing the scenario we propose, and discuss the constraints imposed on it from considerations of the relic density.  In \Sec{sec:benchmarks} we discuss the masses and
couplings of the model in a convenient limit, and present several benchmark points.  \Sec{sec:conclusions} contains our conclusions.

\begin{figure}
\includegraphics[scale=0.25]{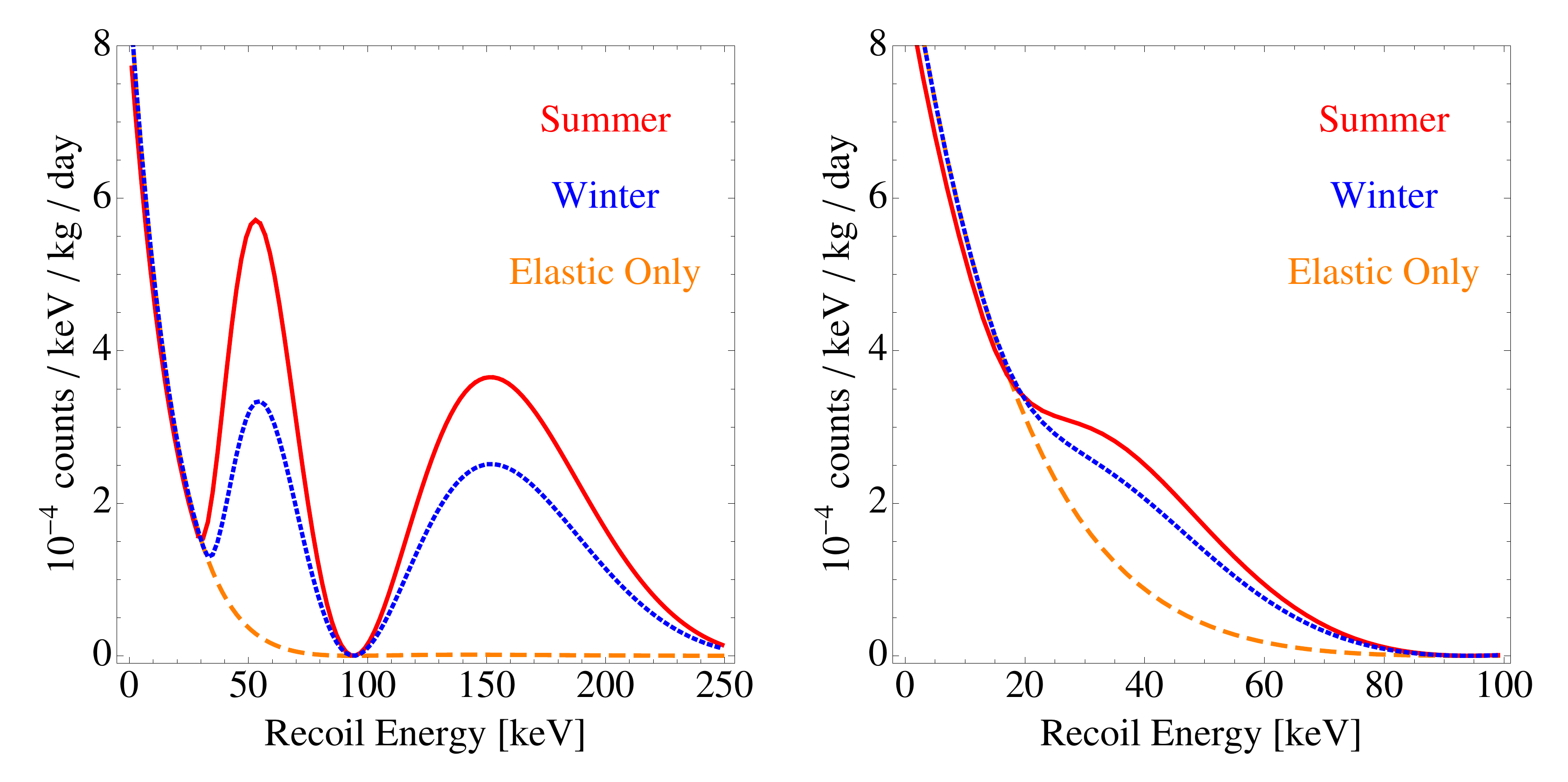}
\caption{\label{fig:semielasticspec} Sample semielastic recoil spectra with Xenon, corresponding to the two benchmark models of Section~\ref{sec:benchmarks}.  We have assumed a local DM density of $\rho=0.3\unit{GeV/cm^3}$, and DM velocities are taken to follow a truncated Maxwell-Boltzmann distribution with $v_0^ {\rm rms}=220$ km/sec and a cutoff of $v_{\rm esc} = 600$ km/sec.  We use the Helm nuclear form factor~\cite{Helm:1956zz}, with the parameterization of Ref.~\cite{Lewin:1995rx}.  The double hump structure on the {\it left} results from the inelastic component of the spectrum overlapping the zero of the form factor at $\sim100$~keV\@.  The summer (winter) spectra correspond to June 2nd (December 2nd).  Both spectra are consistent with current limits \cite{CDMS,Aprile:2010um} and predict $\sim$ 20 events in the first year of XENON100 data (within $8.7\leftrightarrow 40~{\rm keV}$).}
\end{figure}

\section{Semi-elastic Scattering}
The elastic scattering of DM with atomic nuclei can be a natural consequence of the mechanism  setting its TeV-scale mass.  We will illustrate this in an especially simple setup by coupling DM and the Higgs to the same singlet field.  While we will work in a supersymmetric framework, our results are easily generalized to other scenarios.  

Consider the superpotential
\begin{equation}
\label{eq:WS}
W = \lambda S H_d\cdot H_u + \eta S \chi \bar \chi,
\end{equation}
where $H_d$ and $H_u$ are the two Higgs doublet fields, $S$ is the NMSSM singlet~\cite{NMSSM}, and $\chi/\bar \chi$, which are oppositely charged under a dark sector gauge symmetry,  will compose our DM candidate.  After electroweak symmetry breaking, $S$ receives a VEV, which generates a supersymmetric mass for the DM fields, $m_\chi = \eta \left< S \right>$.  Taking DM to be a scalar component of $\chi/\bar\chi$ (we will justify this assumption in the next section), we see that the $F$-term potential includes a direct coupling between DM and the Higgs, $|F_S|^2 \supset \chi \bar \chi H_u^* H_d^* + \mathrm{h.c.}$  This coupling allows the Higgs to mediate the elastic scattering of DM against nuclei.  A second contribution to elastic scattering is mediated by the singlet $S$, which mixes with the Higgs after electroweak symmetry breaking.

At the same time, other dark sector interactions can lead to an inelastic component of scattering.  We consider the class of models where DM is charged under a GeV scale dark sector, with a $U(1)_d$ gauge factor that kinetically mixes with hypercharge,
\be
\cL \supset \frac{\epsilon}{2} B^{\mu \nu} b_{\mu \nu}+ g_d \, b^\mu (\chi_0 \partial_\mu \chi_1 - \chi_1 \partial_\mu \chi_0).
\ee
Here $\chi_{0,1}$ are the two real scalar components of DM separated by a small mass splitting of order $\delta \sim 100$~keV, $B_{\mu}$ and $b_{\mu}$ are the hypercharge and $U(1)_d$ gauge fields, and $\epsilon$ parameterizes the size of the kinetic mixing.  The mass splitting can be generated by a higher dimension operator \cite{Cheung:2009qd}, or radiatively by the breaking of a non-Abelian dark sector gauge symmetry \cite{Baumgart:2009tn}.  Through kinetic mixing, the dark sector photon, $b_{\mu}$, acquires $\epsilon$-suppressed couplings to quarks and thereby mediates inelastic scattering between DM and nuclei.  This scattering will take place along with the elastic scattering described before, realizing a scenario we term {\it semielastic} scattering~\footnote{For a different model that includes inelastic scattering with a subdominant component of elastic scattering, we refer the reader to Ref.~\cite{Lisanti:2009am}.}. 

\begin{figure}
\includegraphics[scale=0.33]{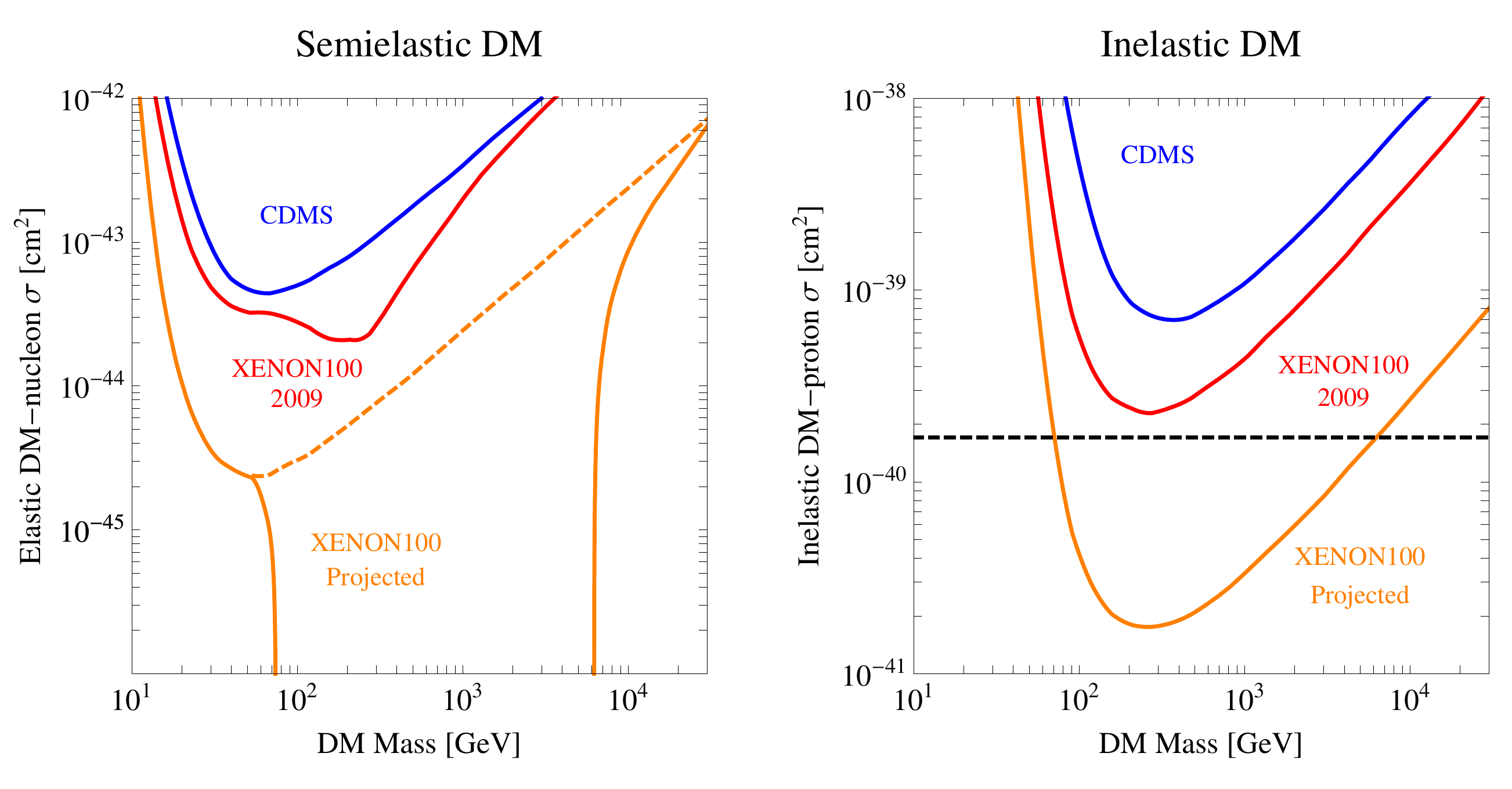}
\caption{\label{fig:limits} The {\it left} panel shows the current 90\% limits, from CDMS~\cite{CDMS} and XENON100~\cite{Aprile:2010um}, on the semielastic scenario in the DM mass - elastic cross-section plane.  We have fixed the DM splitting, $\delta = 140$~keV, and inelastic cross-section, $\sigma_I=1.7\times10^{-40}\unit{cm}^2$ per proton, to match the second benchmark of Section~\ref{sec:benchmarks}.  We have also included the projected limit from XENON100 after one year of data, assuming zero background, a raw exposure of 6000 kg$\times$days, an efficiency of 38\%, and a nuclear recoil energy range of 8.7 to 40 keV\@.  The prominent dip in cross-section indicates the range of masses where the model will be visible from inelastic scattering alone (the dashed orange curve shows the limit without inelastic scattering).  This can be seen on the {\it right} panel which shows the current and projected limits on inelastic scattering only, with $\delta = 140$~keV\@.  The black (dashed) horizontal line on the right corresponds to the inelastic cross-section assumed on the left.
}
\end{figure}

Phenomenologically, semielastic DM can be parameterized by the 4 parameters, $m_{\chi_0}$, $\delta$, $\sigma_{E}$, and $\sigma_{I}$.  The elastic scattering ($\sigma_{E}$) dominates at low nuclear recoil energy while the inelastic scattering ($\sigma_{I}$)  dominates at higher recoil energy. Examples of such a spectrum are shown in \Fig{fig:semielasticspec}.  The unique spectral shape changes the constraints and reach, in the $(\sigma_{E}, m_{\chi_0})$ plane, as shown in \Fig{fig:limits}. We do not attempt to fit the possible DAMA signal~\cite{Bernabei:2010mq}.

\label{sec:semielastic}
\section{Model}
\label{sec:model}
We now provide an explicit realization of the scenario described above.  We take as 
our starting point the NMSSM, where a 
singlet  superfield $S$ couples to the two Higgs multiplets of the MSSM\@.
To this, we add an additional coupling of the singlet to DM, as in Eq.~\ref{eq:WS}, and a $\U(1)_d$ dark sector along the lines of Refs.~\cite{Cheung:2009qd, OtherU1}.

In detail, we will be concerned with the following terms in the superpotential,
\be
\label{eq:superp}
W\supset\lambda S H_d\cdot H_u+\eta S \chi\bar\chi +\frac{1}{3}\kappa S^3+\rho N R \bar R+\frac{1}{\Lambda} \chi^2 \bar R^2
\ee
where $H_d$ and $H_u$ are the two Higgs doublet fields, $S$ is the NMSSM singlet, $\chi$ and $\bar \chi$ will compose our DM candidate,
$R$ and $\bar R$ are GeV-scale dark sector Higgs fields, and $N$ is a  GeV-scale singlet whose presence insures that all dark sector fields receive a tree-level mass~\cite{Cheung:2009qd}.
We will see below that the higher-dimension operator, suppressed by $\Lambda \gtrsim 10$ TeV, will generate a small DM mass splitting.

We assign $\chi$ and $R$ ($\bar \chi$ and $\bar R$ ) charge 1 (-1) under a dark $\U(1)_d$ gauge group.
Furthermore, we assume the presence of supersymmetric
kinetic mixing,
\be
{\cal L}\supset-\frac{\epsilon}{2}\int d^2\theta \, W_Y W_b,
\ee
for $W_Y$ and $W_b$, the hypercharge and dark supersymmetric field strengths, where $\epsilon \sim 10^{-4}\leftrightarrow 10^{-5}$ is naturally generated, at one loop, by integrating out physics at higher energy scales.  Expanding in components, the kinetic mixing includes D-term mixing, which generates an effective Fayet-Iliopoulos D-term in the hidden sector at the GeV scale \cite{Baumgart:2009tn, Cheung:2009qd}.

Upon minimizing the dark sector potential one finds that $R_c$ develops a VEV $\langle R_c\rangle \equiv v_r\sim$ GeV
which Higgses the dark photon, giving it a mass $m_{\gamma_d}=g_d \, v_r$.  The other light dark sector states also live at the GeV-scale.  Finally, we note that there can be ${\cal O}(1)$ corrections to their masses coming from SM
SUSY breaking, which is communicated to the dark sector through gauge interactions with $\chi$ acting as a messenger.  These corrections have been included in our benchmark spectra of section~\ref{sec:benchmarks}, although they have no qualitative effect on the phenomenology.

We now consider the spectrum of the $\chi$ multiplet, which will contain DM\@.  After $S$ gets a VEV, there are two nearly degenerate fermionic states with masses $\sim \eta v_s / \sqrt{2}$ (these states are split a small amount by the higher-dimension operator).  Meanwhile, the scalar components are split from their supersymmetric masses by $\left<F_S\right>$, and under the assumption that $\chi$'s dominant source of SUSY breaking is communicated by $S$, we can neglect additional soft terms.
The four scalar degrees 
of freedom divide into two pairs with masses above and below the fermions, separated by a large weak-scale splitting,
\be
\label{eq:scalardmmass}
m^2=\eta\left[\frac{v_s^2}{2}(\eta\pm\kappa)\mp\frac{\lambda}{4}v_{EW}^2\sin(2\beta)\right],
\ee
where $v_s$ is the singlet VEV\@.  Within each scalar pair there is a smaller splitting
\be
\label{eq:delta}
\delta m^2=\sqrt{2}\eta\frac{v_sv_r^2} {\Lambda}
\ee
where $v_r$ is the VEV of the dark Higgs.  In what follows, we will label the scalar 
mass eigenstates $\chi_i$ for $i:0\rightarrow 3$ in order of ascending mass.  The lightest state, $\chi_0$,
will serve as our DM candidate, and $\delta m=m_{\chi_1}-m_{\chi_0}\sim100$~keV is a consequence of Eq.~\ref{eq:delta}.

\label{sec:relic}
Now, demanding that this model reproduce the observed relic abundance of dark matter places strong
constraints on the different couplings and VEVs.  DM can annihilate via three competitive channels: (1) to Higgses and singlets, (2) via dark sector gauge interactions, and (3) to dark sector Higgses through the higher-dimension contact interaction of Eq.~\ref{eq:superp}.  DM has the observed relic density if these channels sum to have the correct annihilation rate for a thermal relic,
\be
\label{eq:relic}
\langle \sigma v\rangle\sim 2.5 \times 10^{-9}\ {\rm GeV}^{-2}.
\ee
We derive constraints on the model by demanding that no individual channel exceed this rate.  To derive these limits, it is sufficient to consider self-annihilations of $\chi_0$.   In general, there is also co-annihilation of $\chi_0$ with $\chi_1$ (and the $\tilde \chi$ fermions for small enough $\left< F_S \right>$), but the following estimates still apply up to $\cO(1)$ corrections.  We do not consider constraints on Sommerfeld enhanced annihilations from cosmology~\cite{Feng:2010zp}.  Our limits are therefore conservative and also applicable to decaying DM models~\cite{JoshDecayDS}. 

First we consider DM annihilations into the SM Higgses.  Assuming these are light compared to $m_\chi$, one finds
their contribution to the annihilation rate to be
\be
\langle\sigma v\rangle\gtrsim\frac{1}{8\pi}\left(\frac{5 m_{\chi_0}}{v_s^2}\right)^2+\frac{\lambda^4}{2\pi m_{\chi_0}^2}
\ee
which implies
\be
v_s\gtrsim 4.5~{\rm TeV}\left({\frac{m_{\chi_0}}{1~{\rm TeV}}}\right)^{1/2}.
\ee
While there are also constraints on $\lambda$, they are far less severe:
\be
\lambda \lesssim 0.4\left(\frac{m_ {\chi_0}}{1~{\rm TeV}}\right)^{1/2}.
\ee

DM can annihilate through dark sector gauge interactions with rate,
\be
\langle \sigma v\rangle\sim\frac{g_d^4}{8\pi m_ {\chi_0} ^2}
\ee
which constrains the dark gauge coupling:
\be
\left(\frac{g_d}{0.5}\right)^2\left(\frac{1~{\rm TeV}}{m_ {\chi_0}}\right)\lesssim 1.
\ee

In addition, the non-renormalizable operator that generates the small DM splitting allows DM to annihilate into pairs of 
dark-Higgses,
\be
\langle \sigma v\rangle\sim\frac{(\delta m_ {\chi_0})^2}{8\pi v_r^4}.
\ee
Therefore, we find a non-trivial constraint relating the mass splitting between the dark matter states and the dark sector breaking scale \cite{JoshDecayDS},
\be
\left(\frac{\delta m_\chi}{100~{\rm keV}}\right)\left(\frac{1~{\rm GeV}}{v_r}\right)^2\lesssim 1
\ee

We conclude this section with a brief discussion of other important constraints on this model.  We note that the excited state $\chi_1$ is long-lived and has a relic density that is constrained by inelastic down-scattering, but its density can be depleted in several ways as discussed by Ref.~\cite{Finkbeiner:2009mi}.  There are a number of additional constraints if one attempts to explain the cosmic ray anomalies through Sommerfeld enhanced annihilations as in Refs.~\cite{ArkaniHamed:2008qn,OtherDS}.  Dark matter can annihilate into hidden sector gauginos which decay to a dangerous amount of SM photons as discussed by \cite{JoshDecayDS}.  This constraint is alleviated if $m_{\rm gravitino} \gtrsim 1$ GeV, or by considering a more elaborate dark sector.  There is tension from astrophysical limits on neutrinos and photons from final state radiation, coming from the galactic center \cite{PhotonNeutrino}.  Alternatively, these astrophysical tensions are alleviated if the cosmic rays are produced by DM decays into the dark sector \cite{JoshDecayDS}.

\section{Benchmarks}
\label{sec:benchmarks}
\begingroup
\squeezetable
\begin{table*}
\caption{\label{tab:benchmarks}Benchmark points for semielastic scattering.  Here $m_s$ is the mass of the lightest CP-even scalar, which tends to dominate the elastic scattering rate.  The first benchmark has a DM mass that can explain the cosmic ray anomalies with annihilations ($m_{\chi_0}\sim 1~{\rm TeV}$), while the second benchmark has a mass appropriate for decaying DM ($m_{\chi_0}\sim2~{\rm TeV}$)~\cite{PhotonNeutrino}.}
\begin{ruledtabular} 
\begin{tabular}{cccccccccccccc}
$\lambda$ & $\eta$ & $\kappa$ & $\tan\beta$ & $A_\kappa$ & $A_\lambda$ & $v_s$ & $g_d$ & $\epsilon$& $m_{\chi_0}$&$\delta m_{\chi}$&$m_s$&$\sigma_E$ (per nucleon) & $\sigma_I$ (per proton)\\
0.025 & 0.20 & 0.0060 & 15 & $-110~{\rm GeV}$ & $300~{\rm GeV}$ & $7~{\rm TeV}$& 0.5 & $2\cdot10^{-4}$&$975~{\rm GeV}$ & $203~{\rm keV}$& $14~{\rm GeV}$&$1.7\cdot 10^{-43}~{\rm cm}^2$&$8.3\cdot 10^{-39}~{\rm cm}^2$\\ 
0.015 & 0.40 & 0.0015 & 10 & $-15~{\rm GeV}$ & $20~{\rm GeV}$ & $8~{\rm TeV}$& 0.6 & $3\cdot10^{-5}$&$2259~{\rm GeV}$ & $140~{\rm keV}$& $12~{\rm GeV}$&$4.3\cdot 10^{-43}~{\rm cm}^2$&$1.7\cdot 10^{-40}~{\rm cm}^2$\\ 
\end{tabular}
\end{ruledtabular}
\end{table*}
\endgroup
It is convenient to consider this model in the various analytically 
tractable limits of the NMSSM\@.
One finds, however, that whether one starts with a small $\kappa$ (the PQ-symmetric limit~\cite{Hall:2004qd}) 
or with small $A$-terms (the R-symmetric limit~\cite{Dobrescu:2000yn}), the requirement of a sizable $v_s$, a stable
EWSB minima, and an elastic recoil spectra visible at current direct detection experiments necessitates 
small $\kappa$ and $\lambda$.  We therefore consider the limit $\kappa,\lambda\rightarrow 0$.

DM scatters elastically by exchanging the three CP-even Higgses, $s$, $h$, and $H$.  With DM at the TeV scale and the three scalar Higgs masses above 100 GeV, we find that it will be difficult to see the elastic scattering at current direct detection experiments.  Things become more interesting if one of these states is light, enhancing the elastic cross-section.  The mostly-singlet scalar, $s$, enjoys suppressed couplings to the electroweak gauge bosons and can be very light without
conflicting with existing LEP limits~\cite{Barate:2003sz}.  We work in the limit where $m_s \lesssim 50~{\rm GeV}$.  One finds
\be
\sigma_{\rm el}\sim 1.7\cdot 10^{-40}~{\rm cm}^2\left(\frac{m_ {\chi_0}}{v_s}\right)^2\left(\frac{g_H\alpha_{H}+ g_h\alpha_{h}}{m_s^2}\right)^2
\ee
where $g_h\sim1$ and $g_H\sim \frac{1}{2}(\tan\beta-\cot\beta)$ parameterize the couplings of $h$ and $H$ to nucleons~\cite{He:2008qm} (we use the nuclear matrix elements of Ref.~\cite{Giedt:2009mr}), and 
$\alpha_h/\alpha_H$ denote the singlet-Higgs mixing angles.  Note that while it seems one can arbitrarily increase $\sigma_E$ via $g_H$ by choosing a large $\tan\beta$, the 
singlet proportionally decouples from $H$, so no such tuning is possible.

We present 2 benchmark points in \Tab{tab:benchmarks}\@.  Both yield the correct relic abundance and a
recoil spectrum visible in one year of XENON100 data.
\section{Discussion}
\label{sec:conclusions}
Here we have studied a mechanism that naturally relates the mass of DM to the scale
of electroweak symmetry breaking in models with a GeV-scale dark sector.
DM scatters against nuclei elastically via a Higgs/singlet, and inelastically through dark gauge boson exchange.  Combined, the spectrum has a unique semielastic shape which can be discovered in upcoming direct detection experiments such as XENON100.

While our primary concern has been the unique recoil spectrum particular to this class of models, we found in \Sec{sec:model} that the parameters are constrained, nontrivially, by the requirement of getting the right relic density.
It would be interesting to further investigate the interplay between these constraints and the modified Higgs phenomenology of the NMSSM\@.

\acknowledgments{We would like to thank Nima Arkani-Hamed and Neal Weiner for 
pointing out the connection between electroweak symmetry breaking and elastic scattering.  
Also, we would like to acknowledge useful 
discussions with Richard Brower, Adam Falkowski, Jiji Fan, Meifeng Lin, Michele Papucci, Jesse Thaler, and Tomer Volansky.  Finally, Richard Saldanha
suggested (with tongue in cheek) the title of this work.  He got what he deserved.

DK acknowledges travel/computing support from the LHC-TI and thanks the Aspen Center for Physics for their hospitality during the final stages 
of this work.  
JTR is supported by an NSF fellowship.
LTW is supported by the NSF under grant PHY-0756966 and the 
DOE under grant DE-FG02-90ER40542.  
}


\begin{thebibliography}{20}
\expandafter\ifx\csname natexlab\endcsname\relax\def\natexlab#1{#1}\fi
\expandafter\ifx\csname bibnamefont\endcsname\relax
  \def\bibnamefont#1{#1}\fi
\expandafter\ifx\csname bibfnamefont\endcsname\relax
  \def\bibfnamefont#1{#1}\fi
\expandafter\ifx\csname citenamefont\endcsname\relax
  \def\citenamefont#1{#1}\fi
\expandafter\ifx\csname url\endcsname\relax
  \def\url#1{\texttt{#1}}\fi
\expandafter\ifx\csname urlprefix\endcsname\relax\def\urlprefix{URL }\fi
\providecommand{\bibinfo}[2]{#2}
\providecommand{\eprint}[2][]{\url{#2}}

\bibitem[{\citenamefont{Tucker-Smith and Weiner}(2001)}]{TuckerSmith:2001hy}
\bibinfo{author}{\bibfnamefont{D.}~\bibnamefont{Tucker-Smith}}
  \bibnamefont{and} \bibinfo{author}{\bibfnamefont{N.}~\bibnamefont{Weiner}},
  \bibinfo{journal}{Phys. Rev.} \textbf{\bibinfo{volume}{D64}},
  \bibinfo{pages}{043502} (\bibinfo{year}{2001}), \eprint{hep-ph/0101138}.

\bibitem[{\citenamefont{Arkani-Hamed et~al.}(2009)\citenamefont{Arkani-Hamed,
  Finkbeiner, Slatyer, and Weiner}}]{ArkaniHamed:2008qn}
\bibinfo{author}{\bibfnamefont{N.}~\bibnamefont{Arkani-Hamed}},
  \bibinfo{author}{\bibfnamefont{D.~P.} \bibnamefont{Finkbeiner}},
  \bibinfo{author}{\bibfnamefont{T.~R.} \bibnamefont{Slatyer}},
  \bibnamefont{and} \bibinfo{author}{\bibfnamefont{N.}~\bibnamefont{Weiner}},
  \bibinfo{journal}{Phys. Rev.} \textbf{\bibinfo{volume}{D79}},
  \bibinfo{pages}{015014} (\bibinfo{year}{2009}), \eprint{0810.0713}.

\bibitem[{\citenamefont{Pospelov et~al.}(2008)\citenamefont{Pospelov, Ritz, and
  Voloshin}}]{Pospelov:2007mp}
\bibinfo{author}{\bibfnamefont{M.}~\bibnamefont{Pospelov}},
  \bibinfo{author}{\bibfnamefont{A.}~\bibnamefont{Ritz}}, \bibnamefont{and}
  \bibinfo{author}{\bibfnamefont{M.~B.} \bibnamefont{Voloshin}},
  \bibinfo{journal}{Phys. Lett.} \textbf{\bibinfo{volume}{B662}},
  \bibinfo{pages}{53} (\bibinfo{year}{2008}), \eprint{0711.4866}.

\bibitem[{\citenamefont{Nomura and Thaler}(2009)}]{Nomura:2008ru}
\bibinfo{author}{\bibfnamefont{Y.}~\bibnamefont{Nomura}} \bibnamefont{and}
  \bibinfo{author}{\bibfnamefont{J.}~\bibnamefont{Thaler}},
  \bibinfo{journal}{Phys. Rev.} \textbf{\bibinfo{volume}{D79}},
  \bibinfo{pages}{075008} (\bibinfo{year}{2009}), \eprint{0810.5397}.

\bibitem[{\citenamefont{Burgess et~al.}(2001)\citenamefont{Burgess, Pospelov,
  and ter Veldhuis}}]{Burgess:2000yq}
\bibinfo{author}{\bibfnamefont{C.~P.} \bibnamefont{Burgess}},
  \bibinfo{author}{\bibfnamefont{M.}~\bibnamefont{Pospelov}}, \bibnamefont{and}
  \bibinfo{author}{\bibfnamefont{T.}~\bibnamefont{ter Veldhuis}},
  \bibinfo{journal}{Nucl. Phys.} \textbf{\bibinfo{volume}{B619}},
  \bibinfo{pages}{709} (\bibinfo{year}{2001}), \eprint{hep-ph/0011335}.

\bibitem[{\citenamefont{March-Russell et~al.}(2008)\citenamefont{March-Russell,
  West, Cumberbatch, and Hooper}}]{MarchRussell:2008yu}
\bibinfo{author}{\bibfnamefont{J.}~\bibnamefont{March-Russell}},
  \bibinfo{author}{\bibfnamefont{S.~M.} \bibnamefont{West}},
  \bibinfo{author}{\bibfnamefont{D.}~\bibnamefont{Cumberbatch}},
  \bibnamefont{and} \bibinfo{author}{\bibfnamefont{D.}~\bibnamefont{Hooper}},
  \bibinfo{journal}{JHEP} \textbf{\bibinfo{volume}{07}}, \bibinfo{pages}{058}
  (\bibinfo{year}{2008}), \eprint{0801.3440}.

\bibitem[{\citenamefont{Arina et~al.}(2010)\citenamefont{Arina, Josse-Michaux,
  and Sahu}}]{Arina:2010an}
\bibinfo{author}{\bibfnamefont{C.}~\bibnamefont{Arina}},
  \bibinfo{author}{\bibfnamefont{F.-X.} \bibnamefont{Josse-Michaux}},
  \bibnamefont{and} \bibinfo{author}{\bibfnamefont{N.}~\bibnamefont{Sahu}}
  (\bibinfo{year}{2010}), \eprint{1004.3953}.

\bibitem[{\citenamefont{Farina et~al.}(2009)\citenamefont{Farina, Pappadopulo,
  and Strumia}}]{Farina:2009ez}
\bibinfo{author}{\bibfnamefont{M.}~\bibnamefont{Farina}},
  \bibinfo{author}{\bibfnamefont{D.}~\bibnamefont{Pappadopulo}},
  \bibnamefont{and} \bibinfo{author}{\bibfnamefont{A.}~\bibnamefont{Strumia}}
  (\bibinfo{year}{2009}), \eprint{0912.5038}.

\bibitem[{\citenamefont{Nilles et~al.}(1983)\citenamefont{Nilles, Srednicki,
  and Wyler}}]{Nilles:1982dy}
\bibinfo{author}{\bibfnamefont{H.~P.} \bibnamefont{Nilles}},
  \bibinfo{author}{\bibfnamefont{M.}~\bibnamefont{Srednicki}},
  \bibnamefont{and} \bibinfo{author}{\bibfnamefont{D.}~\bibnamefont{Wyler}},
  \bibinfo{journal}{Phys. Lett.} \textbf{\bibinfo{volume}{B120}},
  \bibinfo{pages}{346} (\bibinfo{year}{1983}).

\bibitem[{\citenamefont{Frere et~al.}(1983)\citenamefont{Frere, Jones, and
  Raby}}]{Frere:1983ag}
\bibinfo{author}{\bibfnamefont{J.~M.} \bibnamefont{Frere}},
  \bibinfo{author}{\bibfnamefont{D.~R.~T.} \bibnamefont{Jones}},
  \bibnamefont{and} \bibinfo{author}{\bibfnamefont{S.}~\bibnamefont{Raby}},
  \bibinfo{journal}{Nucl. Phys.} \textbf{\bibinfo{volume}{B222}},
  \bibinfo{pages}{11} (\bibinfo{year}{1983}).

\bibitem[{\citenamefont{Derendinger and Savoy}(1984)}]{Derendinger:1983bz}
\bibinfo{author}{\bibfnamefont{J.~P.} \bibnamefont{Derendinger}}
  \bibnamefont{and} \bibinfo{author}{\bibfnamefont{C.~A.} \bibnamefont{Savoy}},
  \bibinfo{journal}{Nucl. Phys.} \textbf{\bibinfo{volume}{B237}},
  \bibinfo{pages}{307} (\bibinfo{year}{1984}).

\bibitem[{\citenamefont{Cheung et~al.}(2009)\citenamefont{Cheung, Ruderman,
  Wang, and Yavin}}]{Cheung:2009qd}
\bibinfo{author}{\bibfnamefont{C.}~\bibnamefont{Cheung}},
  \bibinfo{author}{\bibfnamefont{J.~T.} \bibnamefont{Ruderman}},
  \bibinfo{author}{\bibfnamefont{L.-T.} \bibnamefont{Wang}}, \bibnamefont{and}
  \bibinfo{author}{\bibfnamefont{I.}~\bibnamefont{Yavin}},
  \bibinfo{journal}{Phys. Rev.} \textbf{\bibinfo{volume}{D80}},
  \bibinfo{pages}{035008} (\bibinfo{year}{2009}), \eprint{0902.3246}.

\bibitem[{\citenamefont{Meade et~al.}(2009)\citenamefont{Meade, Papucci, and
  Volansky}}]{Meade:2009rb}
\bibinfo{author}{\bibfnamefont{P.}~\bibnamefont{Meade}},
  \bibinfo{author}{\bibfnamefont{M.}~\bibnamefont{Papucci}}, \bibnamefont{and}
  \bibinfo{author}{\bibfnamefont{T.}~\bibnamefont{Volansky}},
  \bibinfo{journal}{JHEP} \textbf{\bibinfo{volume}{12}}, \bibinfo{pages}{052}
  (\bibinfo{year}{2009}), \eprint{0901.2925}.

\bibitem[{\citenamefont{Finkbeiner et~al.}(2009)\citenamefont{Finkbeiner,
  Slatyer, Weiner, and Yavin}}]{Finkbeiner:2009mi}
\bibinfo{author}{\bibfnamefont{D.~P.} \bibnamefont{Finkbeiner}},
  \bibinfo{author}{\bibfnamefont{T.~R.} \bibnamefont{Slatyer}},
  \bibinfo{author}{\bibfnamefont{N.}~\bibnamefont{Weiner}}, \bibnamefont{and}
  \bibinfo{author}{\bibfnamefont{I.}~\bibnamefont{Yavin}},
  \bibinfo{journal}{JCAP} \textbf{\bibinfo{volume}{0909}}, \bibinfo{pages}{037}
  (\bibinfo{year}{2009}), \eprint{0903.1037}.

\bibitem[{\citenamefont{Ruderman and Volansky}(2010)}]{Ruderman:2009tj}
\bibinfo{author}{\bibfnamefont{J.~T.} \bibnamefont{Ruderman}} \bibnamefont{and}
  \bibinfo{author}{\bibfnamefont{T.}~\bibnamefont{Volansky}},
  \bibinfo{journal}{JHEP} \textbf{\bibinfo{volume}{02}}, \bibinfo{pages}{024}
  (\bibinfo{year}{2010}), \eprint{0908.1570}.

\bibitem[{\citenamefont{Hall and Watari}(2004)}]{Hall:2004qd}
\bibinfo{author}{\bibfnamefont{L.~J.} \bibnamefont{Hall}} \bibnamefont{and}
  \bibinfo{author}{\bibfnamefont{T.}~\bibnamefont{Watari}},
  \bibinfo{journal}{Phys. Rev.} \textbf{\bibinfo{volume}{D70}},
  \bibinfo{pages}{115001} (\bibinfo{year}{2004}), \eprint{hep-ph/0405109}.

\bibitem[{\citenamefont{Dobrescu and Matchev}(2000)}]{Dobrescu:2000yn}
\bibinfo{author}{\bibfnamefont{B.~A.} \bibnamefont{Dobrescu}} \bibnamefont{and}
  \bibinfo{author}{\bibfnamefont{K.~T.} \bibnamefont{Matchev}},
  \bibinfo{journal}{JHEP} \textbf{\bibinfo{volume}{09}}, \bibinfo{pages}{031}
  (\bibinfo{year}{2000}), \eprint{hep-ph/0008192}.

\bibitem[{\citenamefont{Barate et~al.}(2003)}]{Barate:2003sz}
\bibinfo{author}{\bibfnamefont{R.}~\bibnamefont{Barate}} \bibnamefont{et~al.}
  (\bibinfo{collaboration}{LEP Working Group for Higgs boson searches}),
  \bibinfo{journal}{Phys. Lett.} \textbf{\bibinfo{volume}{B565}},
  \bibinfo{pages}{61} (\bibinfo{year}{2003}), \eprint{hep-ex/0306033}.

\bibitem[{\citenamefont{Giedt et~al.}(2009)\citenamefont{Giedt, Thomas, and
  Young}}]{Giedt:2009mr}
\bibinfo{author}{\bibfnamefont{J.}~\bibnamefont{Giedt}},
  \bibinfo{author}{\bibfnamefont{A.~W.} \bibnamefont{Thomas}},
  \bibnamefont{and} \bibinfo{author}{\bibfnamefont{R.~D.} \bibnamefont{Young}},
  \bibinfo{journal}{Phys. Rev. Lett.} \textbf{\bibinfo{volume}{103}},
  \bibinfo{pages}{201802} (\bibinfo{year}{2009}), \eprint{0907.4177}.

\bibitem[{\citenamefont{He et~al.}(2009)\citenamefont{He, Li, Li, Tandean, and
  Tsai}}]{He:2008qm}
\bibinfo{author}{\bibfnamefont{X.-G.} \bibnamefont{He}},
  \bibinfo{author}{\bibfnamefont{T.}~\bibnamefont{Li}},
  \bibinfo{author}{\bibfnamefont{X.-Q.} \bibnamefont{Li}},
  \bibinfo{author}{\bibfnamefont{J.}~\bibnamefont{Tandean}}, \bibnamefont{and}
  \bibinfo{author}{\bibfnamefont{H.-C.} \bibnamefont{Tsai}},
  \bibinfo{journal}{Phys. Rev.} \textbf{\bibinfo{volume}{D79}},
  \bibinfo{pages}{023521} (\bibinfo{year}{2009}), \eprint{0811.0658}.

\end{thebibliography}


\begin{thebibliography}{9}


\bibitem{Cosmics}
  O.~Adriani {\it et al.},
  Nature {\bf 458}, 607 (2009), 0810.4995.
    F.~Aharonian {\it et al.},
  Phys.\ Rev.\ Lett.\  {\bf 101}, 261104 (2008), 0811.3894.
    F.~Aharonian {\it et al.},
  Astron.\ Astrophys.\  {\bf 508}, 561 (2009), 0905.0105.
    A.~A.~Abdo {\it et al.},
  Phys.\ Rev.\ Lett.\  {\bf 102}, 181101 (2009), 0905.0025.

\bibitem{ArkaniHamed:2008qn}
  N.~Arkani-Hamed {\it et al.},
  Phys.\ Rev.\  D {\bf 79}, 015014 (2009), 0810.0713.
  
\bibitem{ArkaniHamed:2008qp}
  N.~Arkani-Hamed and N.~Weiner,
  JHEP {\bf 0812}, 104 (2008), 0810.0714.
  
\bibitem{OtherDS}
  M.~Pospelov {\it et al.},
  Phys.\ Lett.\  B {\bf 662}, 53 (2008), 0711.4866.
  Y.~Nomura and J.~Thaler,
  Phys.\ Rev.\  D {\bf 79}, 075008 (2009), 0810.5397.
  
  \bibitem{Baumgart:2009tn}
  M.~Baumgart {\it et al.},
  JHEP {\bf 0904}, 014 (2009), 0901.0283.
  
\bibitem{Cheung:2009qd}
  C.~Cheung {\it et al.},
  Phys.\ Rev.\  D {\bf 80}, 035008 (2009), 0902.3246.

\bibitem{OtherDecayDS}
  X.~Chen,
  JCAP {\bf 0909}, 029 (2009), 0902.0008.
    J.~Mardon {\it et al.},
  Phys.\ Rev.\  D {\bf 80}, 035013 (2009), 0905.3749.

\bibitem{JoshDecayDS}
  J.~T.~Ruderman and T.~Volansky,
  JHEP {\bf 1002}, 024 (2010), 0908.1570.
  J.~T.~Ruderman and T.~Volansky,
  arXiv: 0907.4373.
 
 \bibitem{Inelastic}
  D.~Tucker-Smith and N.~Weiner,
  Phys.\ Rev.\  D {\bf 64}, 043502 (2001), hep-ph/0101138.
  D.~Tucker-Smith and N.~Weiner,
  Phys.\ Rev.\  D {\bf 72}, 063509 (2005), 0402065.
  S.~Chang {\it et al.},
  Phys.\ Rev.\  D {\bf 79}, 043513 (2009), 0807.2250.
  
\bibitem{Bernabei:2010mq}
  R.~Bernabei {\it et al.},
  Eur.\ Phys.\ J.\  C {\bf 67}, 39 (2010), 1002.1028.
 
 \bibitem{HiggsSDM}
  J.~March-Russell {\it et al.},
  JHEP {\bf 0807}, 058 (2008), 0801.3440.
    D.~E.~Kaplan {\it et al.},
  Phys.\ Rev.\  D {\bf 79}, 115016 (2009), 0901.4117.
  
\bibitem{Helm:1956zz}
  R.~H.~Helm,
  Phys.\ Rev.\  {\bf 104}, 1466 (1956).
  
\bibitem{Lewin:1995rx}
  J.~D.~Lewin and P.~F.~Smith,
  Astropart.\ Phys.\  {\bf 6}, 87 (1996).

\bibitem{CDMS}
  D.~S.~Akerib {\it et al.} 
  Phys.\ Rev.\ Lett.\  {\bf 93}, 211301 (2004), astro-ph/0405033.
    D.~S.~Akerib {\it et al.} 
  Phys.\ Rev.\ Lett.\  {\bf 96}, 011302 (2006), astro-ph/0509259.
    Z.~Ahmed {\it et al.} 
  Phys.\ Rev.\ Lett.\  {\bf 102}, 011301 (2009), 0802.3530.
    Z.~Ahmed {\it et al.},
  arXiv:0912.3592.
  
\bibitem{Aprile:2010um}
  E.~Aprile {\it et al.},
  arXiv:1005.0380.


  


\bibitem{NMSSM}
  H.~P.~Nilles {\it et al.},
  Phys.\ Lett.\  B {\bf 124}, 337 (1983).
    J.~M.~Frere {\it et al.},
  Nucl.\ Phys.\  B {\bf 222}, 11 (1983).
   J.~P.~Derendinger and C.~A.~Savoy,
  Nucl.\ Phys.\  B {\bf 237}, 307 (1984).

\bibitem{Lisanti:2009am}
  M.~Lisanti and J.~G.~Wacker,
  arXiv:0911.4483.
  
  \bibitem{OtherU1}
  A.~Katz and R.~Sundrum,
  JHEP {\bf 0906}, 003 (2009), 0902.3271.
  D.~E.~Morrissey {\it et al.},
  JHEP {\bf 0907}, 050 (2009), 0904.2567.

\bibitem{Feng:2010zp}
  J.~L.~Feng {\it et al.}, 1005.4678.
  M.~R.~Buckley and P.~J.~Fox,
  Phys.\ Rev.\  D {\bf 81}, 083522 (2010), 0911.3898.

\bibitem{Finkbeiner:2009mi}
  D.~P.~Finkbeiner {\it et al.},
  JCAP {\bf 0909}, 037 (2009), 0903.1037.

\bibitem{PhotonNeutrino}
    P.~Meade {\it et al.},
  Nucl.\ Phys.\  B {\bf 831}, 178 (2010), 0905.0480.

    \bibitem{Hall:2004qd}
  L.~J.~Hall and T.~Watari,
  Phys.\ Rev.\  D {\bf 70}, 115001 (2004), hep-ph/0405109.

\bibitem{Dobrescu:2000yn}
  B.~A.~Dobrescu and K.~T.~Matchev,
  JHEP {\bf 0009}, 031 (2000), hep-ph/0008192.

\bibitem{Barate:2003sz}
  R.~Barate {\it et al.}
  Phys.\ Lett.\  B {\bf 565}, 61 (2003), hep-ex/0306033.

\bibitem{He:2008qm}
  X.~G.~He {\it et al.},
  Phys.\ Rev.\  D {\bf 79}, 023521 (2009), 0811.0658.
  
\bibitem{Giedt:2009mr}
  J.~Giedt {\it et al.},
  Phys.\ Rev.\ Lett.\  {\bf 103}, 201802 (2009), 0907.4177.

\end{thebibliography}
\end{document}